\documentclass[a4paper]{jpconf}
\usepackage{graphicx}
\begin{document}
\title{Multi-messenger cosmology of new physics}

\author{Maxim Yu. Khlopov}

\address{Institute of Physics, Southern Federal University, Rostov on Don and National Research Nuclear University "MEPHI", Moscow, Russia\\ Université de Paris, CNRS, Astroparticule et Cosmologie, F-75013 Paris, France}

\ead{khlopov@apc.univ-paris7.fr}

\begin{abstract}
The observational evidence for the inflationary cosmology with baryosynthesis and dark matter/energy can be viewed as the messenger for new physics, which governed the Universe origin, evolution and structure. To specify the physics beyond the Standard model (BSM), underlying the modern cosmological paradigm additional model dependent messengers are proposed, involving multi-component and composite dark matter, meta-stable particles, primordial black holes and antimatter domains in baryon asymmetrical Universe. 
\end{abstract}

\section{Introduction}
The data of precision cosmology strongly tighten deviations from the predictions of inflationary model with baryosynthesis and dark matter/energy \cite{Lindebook,Kolbbook,Rubakovbook1,Rubakovbook2,book,newBook}. All these basic elements of the modern cosmological paradigm find their physical nature in predictions of BSM models, which in their turn involve cosmological probes for their test \cite{book,newBook,4,DMRev}. 

In the context of the modern cosmological paradigm we may consider the observed structure and evolution of the Universe as the messenger of BSM physics, which needs additional model dependent signatures to be specified. Such model dependent cosmological signatures reflect the fundamental structure and symmetry breaking pattern of the BSM model \cite{4,DMRev,ijmpd19} and can be viewed as multi-messenger cosmological probes for new physics. Here we briefly review some examples of these probes. They involve new stable and meta-stable particles, multi-component dark matter, composite dark matter and dark atoms, primordial black holes and primordial nonlinear structures, as well as antimatter stars in the baryon asymmetrical Universe as a profound signature of strongly nonhomogeneous baryosynthesis. Such consequences are not inevitable predictions of BSM models, but reminding Ya.B.Zeldovich one can say that "even if the probability for these phenomena is very low, the expectation value of their discovery would be very high".

In general, effects of new physics with energy scale $V$ appear with full strength at high energies $E \ge V$. At these energies new particles with the mass characterized by the scale $V$ can be copiously produced, as well as their exchange is not suppressed. At smaller energies $E < V$ these particles can be produced in virtual states and their effects are suppressed by some power of $(E/V)$. 

For high energy scale $V$, cosmology, predicting the stages of early Universe with very high energy density, becomes natural laboratory of new physics. Its observable signatures require some messengers, which retain information on the processes in the very early Universe and provide their confrontation with the astrophysical data on the phenomena, taking place at much later stage of cosmological evolution. It implies sufficiently long-living particles and objects, surviving sufficiently long period after their creation. From the view point of particle theory such particles and objects reflect the fundamental symmetry of BSM model and mechanisms of its symmetry breaking, making cosmological messengers tracers of the fundamental symmetry of microworld. Here we briefly discuss some examples of the messengers of new physics. 

\section{Probes for dark matter physics}
Nonbaryonic dark matter, dominating in the matter content of the modern Universe, is associated with the new stable form of the nonrelativistic matter. It should be nonluminous and must decouple from plasma and radiation before the beginning of the matter dominated stage. The first condition follows from the "darkness" of this form of matter. The second comes from the condition that dark matter provides effective development of gravitational instability in the beginning of matter dominated stage before recombination of hydrogen (see e.g. \cite{4,DMRev,ijmpd19} for reviews and references). The simplest theoretical possibility to satisfy these conditions is to assume the existence of stable neutral Weakly Interacting Massive particles (WIMP). 
\subsection{From WIMP miracle to Dark Matter reality?}
\subsubsection{WIMP miracle}
The attractive feature of the WIMP dark matter candidates was their miraculous property to explain the observed dark matter density by primordial gas of stable particles with mass of the order of several hundred GeV with annihilation cross section of the order of the ordinary weak interaction. These conditions naturally lead to the predicted abundance corresponding to the measured density of dark matter.

Strong theoretical support for WIMPs came from predictions of  stable lightest supersymmetric (SUSY) neutral particles with mass and annihilation cross section, corresponding to the desired WIMP parameter range. The advantage of supersymmetry with SUSY scale within 1 TeV was its principle possibility to solve the problems of Standard model related with divergence of Higgs boson mass and origin of the scale of the electroweak symmetry breaking. The expected discovery of supersymmetric partners of ordinary quarks, leptons and gauge bosons with the mass in the range 100 GeV-1 TeV, was the challenge for experimental search at the LHC.

However, the results of the direct WIMP search in underground experiments are controversial, as well as there is no positive results of SUSY particle searches at the LHC in the indicated mass range, It stimulates the substantial extension of the list of possible dark matter particle candidates.
\subsubsection{Non-WIMP Dark Matter candidates}
Stability of dark matter implies stability of its constituents, which involves new stable or very long-living particles, predicted by BSM models. It assumes extension of the symmetry of the Standard model, which leads to new conserved charges, corresponding the the new additional symmetry. The lightest particle, which possess new charge is stable, if the charge is strictly conserved. 

There are several strongly motivated extensions of the Standard model, predicting various types of dark matter candidates (see e.g. \cite{4} for review and references):
\begin{itemize}
    \item Sterile neutrinos, having no ordinary weak interaction and involved in the see-saw mechanism of neutrino mass generation;
    \item axion, a pseudo Nambu-Goldstone boson related with the Peccei-Quinn solution of the problem of strong CP violation in QCD;
    \item mirror or shadow matter, restoring equivalence of left- and right- handed coordinate systems. Being in the same space-time with the ordinary matter they have gravitational interaction and can also interact with matter due to strongly suppressed kinetic mixing of neutral bososns, like mixing of ordinary and mirror photons.  
    \item gravitino, SUSY partner of graviton in Supergravity. By construction gravitino has super-weak semi-gravitational interaction. At very high sub-Planckean SUSY energy scale it can be also superheavy
\end{itemize}
These extensions of the Standard model lead to non-WIMP dark matter candidates. Sterile neutrinos, mirror or shadow particles or gravitino are superWIMPs with superweak interaction with matter, while axions have a very mall mass, but still play the role of Cold Dark Matter. The list of these candidates can be extended by neutral stable particles originated by any extension of the group of the SM symmetry SU(3) x SU(2) x U(1) by any additional strict symmetry group $G$. In particular, new stable colored objects that possess the corresponding new conserved charge can form Strongly Interacting Massive Particles (SIMP). 
\subsubsection{Multicomponent dark matter}
The motivation for existence of various dark matter particle candidates come from different solutions for the internal problems of SM. It makes possible their co-existence and can lead to multicomponent dark matter scenarios.

In such scenarios dark matter can represent mixture of primordial particles with different properties, like mixture of Hot and Cold Dark matter. Another possibility is co-existence of absolutely stable and metastable particles. The latter can lead to observable effects of deviations from the Standard cosmological scenario. 

To be of cosmological significance metastable particles with the mass $m$ must be sufficiently long living. Their lifetime $\tau$ should be much larger than $m_{Pl}/m^2$. Then they retain in the Big Bang Universe at $T \ll m$ and their presence can lead to observable signatures.
\subsection{Cosmoarcheology of new physics}
The set of astrophysical data puts constraints on any new forms of matter present in the Universe at various periods of cosmological evolution. The very fact of their presence means that they contribute to the total density and such contribution is restricted by the measured density, or effects of their presence in the period of Big Bang Nucleosynthesis or Large Scale Structure formation.

Metastable particle with lifetime $\tau$ exceeding the age of the Universe $t_U$ should contribute the modern dark matter density as decaying dark matter component. If leptons, quarks, gluons or photons are among the decay products, their contributions in the cosmic ray fluxes can provide constraints on the lifetime, branching ratios and abundance of metastable particles. 

Metastable particles with lifetime $\tau < t_U$ cannot be considered as the candidates for the modern dark matter, but their presence in the period of structure formation can lead to unstable dark matter (UDM) scenarios, which are severely constrained by the condition of the effective growth of density fluctuations, which can be strongly suppressed after decays, if UDM dominates in the period of large scale structure formation.

\begin{figure}
\begin{center}
\includegraphics[width=30pc]{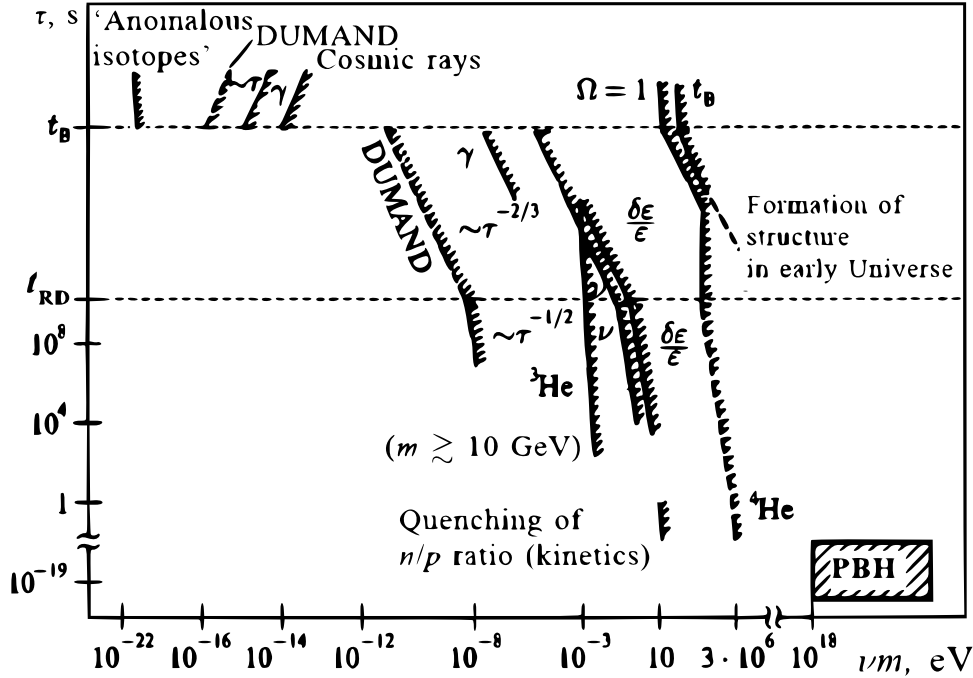}
\end{center}
\caption{\label{ketov}Constraints on the lifetime $\tau$  and product of relative concentration $\nu$ and mass $m$ of metastable particles \cite{ketovSym}. }
\end{figure}

The sensitivity of astrophysical data to the presence and decays of metastable particles is illustrated on the Fig. \ref{ketov}. It strongly depends on the contribution of the decaying particles into the total density and on the possibility of decay products to influence the observable features of the CMB spectrum, light element abundance or cosmic neutrino, gamma ray or cosmic ray fluxes. This sensitivity strongly increases, if decay products influence observable features of subdominant component, which is baryonic matter at the radiation dominated stage and radiation at the matter dominated stage. Such sensitive probes assume  specific decay channels and are strongly model dependent. Contribution to the total density of the Universe at various periods of cosmological evolution avoids such specific model dependence, but on this reason is much less sensitive to the presence of new particles in the Universe.

\subsection{Composite dark matter}
\subsubsection{Problem of stable charged particles}
BSM models try to avoid predictions of stable electrically charged particles. Positively charged stable particles should bind with electrons and form anomalous isotopes of chemical elements. The constraints on the presence of such anomalous isotopes in the terrestrial matter put severe constraint on their abundance.  Only superheavy subPlanckean Charged Massive particles (CHAMP) can avoid these constraints due to very small number density and rapid diffusion to the center of Earth, strongly reducing their abundance in the sea and terrestrial layers near the surface.

Similar to baryonic matter, charged stable particles may be hidden in neutral atomic states and play the role of dark matter. The only condition is to avoid overproduction of anomalous isotopes in this case. The main problem is that in the expanding Universe recombination of electrically charged particles is never complete and freezing out of free charged particles is inevitable. Free +1 charge particles form anomalous hydrogen, severely constrained by the experimental data. Free -1 charged particles $E^-$ form +1 charged ion ($E He$) with primordial helium nuclei, as soon as they are produced in the Big Bang Nucleosynthesis. Similar problems arise for all positively charged stable particles and negatively charged particles with charge $-2n-1$. It leaves only $-2n$ charged particles as possible constituents of dark atoms.

\subsubsection{Multicharged stable particles}
Multicharged particles may be composite or elementary. The example of composite -2 charged particles give models with new stable $U$-type quark. They predict existence of stable $\Delta^{--}$ - like state ($\bar U \bar U \bar U$). It's positively charged antiparticle ($UUU$) can bind with electrons in anomalous helium and special mechanisms are needed to suppress their abundance. Such mechanisms may naturally appear, if the ($\bar U \bar U \bar U$) excess over ($UUU$) is generated similar to baryon excess in baryon asymmetrical Universe.

The balance between excess of new particles and baryon asymmetry can be established by sphaleron transitions, if new particles possess electroweak SU(2) charges. Such balance with proper (negative) sign of the excessive new particles takes place in Walking Technicolor (WTC) models, predicting technibaryons composed of techniquarks and elementary technileptons. The absolute values of electric charges of technibaryons and technileptons are free parameter of the model. The only condition for the charge assignment is the cancellation of anomalies that fixes the relationship between the charges of technileptons and technibaryons, while the absolute value of these charges depends on the free parameter of this model. New stable charged techniparticles may be technibaryons, if technibaryon charge is conserved, technileptons, if technilepton charge is conserved, or both, if the both charges are conserved. In the latter case two-component techniparticle dark matter scenario is possible. Both technibaryons and technileptons look like elementary leptons at energies below WTC confinement.

\subsubsection{Dark atoms of dark matter}
Independent of the mechanism of baryon excess generation, sphaleron transitions establish equilibrium between baryon excess and excess of charged techniparticles. Choice of reasonable parameters of the model provides excess of even negatively charged stable techniparticles, which provides their explanation of the observed dark matter density for the masses of the order of 1 TeV.

After Big Bang Nucleosynthesis these excessive $-2n$ charged techniparticles bind with $n$ helium nuclei in dark atoms. $O^{--}$ with charge -2 form $O$He atoms - Bohr like systems with $O^{--}$ leptonic core and strongly interacting helium shell. The Bohr radius in $O$He atom is equal to the size of He. The lack of usual approximations of atomic physics (small size of nuclear interacting nucleus relative to Bohr orbit and electronic shell with electroweak interaction, supporting perturbation methods of calculations) makes proper quantum mechanical treatment of $O$He interaction with matter a very complicated and still unresolved problem.
\subsubsection{Multimessenger probes for dark atoms}
Cosmological scenario of dark atom evolution leads to Warmer than Cold dark matter scenario of structure formation. Owing to low number density of nuclei $O$He gas decouples from plasma and radiation before the beginning of matter dominated stage and supports growth of density fluctuations with spectrum with slightly suppressed short wave part as compared with the standard Cold dark matter scenario. 

In spite of its strong interaction with matter ($\sigma \approx 2 10^{-25}$ cm$^2$), only sufficiently dense matter objects of the size $R$ with density $$\rho > \frac{1}{\sigma R m_p},$$ where $m_p$ is the mass of proton, are opaque for $O$He, while the average matter density makes the Galaxy transparent for it. $O$He gas in the Galaxy is collisionless, but in the region of the Galaxy center, where $O$He density is higher rare $O$He collisions can lead to $O$He excitations. De-excitation of $O$He, excited in collisions, by emission of electron-positron pairs can provide explanation for the excess of positron annihilation line radiation from the galactic bulge, observed by INTEGRAL. Such explanation implies the mass of $O^{--}$ in the narrow window near 1.25 TeV, challenging the search of such stable double charged particles at the LHC.

Due to strong interaction with matter cosmic $O$He is slowed down in the terrestrial matter and cannot be detected in underground experiments by effects of nuclear recoil, used for direct WIMP searches. However, annual modulation in low energy binding of $O$He with intermediate mass nuclei, like sodium, can explain the positive results of DAMA/NaI and DAMA/LIBRA experiments with their puzzling contradictions with negative results of direct WIMP searches.

Created after helium production in the Big Bang Nucleosynthesis $O$He can catalyze pregalactic production of heavier nuclei, like carbon or oxygen. Captured by stars $O$He can play interesting but still unexplored role in stellar evolution. Liberated in stellar interiors and accelerated at Supernova explosions multiple charged dark atom constituents can form high energy flux of exotic multiple charged leptonic component that can lead to specific type of atmospheric showers in LHAASO experiment.

\section{Messengers of very early Universe} 
Together with baryon asymmetry or primordial gas of dark matter particles physics of very early Universe can provide many other model dependent observable tracers. Second order phase transitions can lead to formation of topological defects like monopoles, strings, walls or many other types of stable or unstable topological defects. Strong first order phase transitions can be the source of gravitational wave background.
These processes can lead to appearance of inhomogeneities in homogeneous and isotropic Universe.

One of the profound signature of strong inhomogeneity of very early Universe is formation of primordial black holes. Their spectrum contains information on the mechanisms of their formation, reflecting the fundamental structure of the particle theory at very high energy scale \cite{4,PBHrev}. 

\subsection{Primordial Black Holes as the tracer of new physics}
\label{PBH}
To form black hole in the expanding Universe, one should stop its expansion within the cosmological horizon \cite{ZN}. It corresponds to nonhomogeneity $\delta = \delta \rho/\rho \sim 1$ in the nearly homogeneous and isotropic Universe with dispersion of small density fluctuations \begin{equation}
\left\langle \delta^2 \right\rangle = \delta_o^2 \ll 1 \label{pbh} \end{equation}. Probability for such a high amplitude fluctuation depends on the equation of state $p= \gamma \epsilon$ (where $p$ is pressure, $\epsilon$ is energy density and $\gamma = 0$ for matter dominance (MD) and $\gamma= 1/3$ for radiation dominated (RD) stage) and is given by \cite{carr75} 
$$W_{PBH} \propto \exp \left(-\frac{\gamma^2}{2
\left\langle \delta^2 \right\rangle}\right).$$
This probability is exponentially suppressed for small amplitude density fluctuations at the RD stage. At MD stage there is no exponential suppression. It makes primordial black holes a sensitive indicator of early MD stages \cite{polnarev,khlopov0}.
\subsubsection{Physics of early MD stages}
Early MD stage may be a consequence of existence of a supermassive metastable particle, dominating in the Universe before decay \cite{PBHrev,polnarev,khlopov0}. If such particles with mass $m$ are created in the Big Bang Universe with frozen out relative abundance $\nu = n_m/n_r$, where $n_m$ and $n_r$ are number densities of considered particles and relativistic species, respectively, at the temperature $T < T_o =\nu m$, corresponding to the period  $t>t_o = m_{Pl}/m^2$ such particles start to dominate in the Universe until their decay at $t=\tau$,
where $\tau$ is the particle lifetime.

Growth of density fluctuations at the MD stage leads to formation of gravitationally bound systems, separated from cosmological expansion. Evolution of these systems can lead to formation of black holes, retaining in the Universe at $t>\tau$, when particles, dominating in the Universe, decay. 

The minimal estimation of the probability of PBH formation is independent on the nature of particles, being determined by direct collapse into black hole of specially homogeneous and isotropic configurations, after they separate from the general expansion. This probability is given by \cite{polnarev}
$$ W_{PBH} \propto \delta_o^{13/2}.$$
If configuration is specially homogeneous and isotropic it contracts within its gravitational radius as soon as it separates from cosmological expansion at $t_1 \approx t_0 \delta_o^{-3/2}$. This mechanism leads to a flat spectrum of PBH masses ranging from $M_{min}=m_{Pl}^2 t_o$ to the maximal mas, determined by the condition that the configuration can separate from expansion and collapse in black hole before particles decay at $t=\tau$.

However, dominant part of configurations don't contract directly into black holes and form gravitationally bound systems, whose evolution strongly depends on the nature of particles, dominating at the MD stage. 

If gas of massive particles within configuration is collisionless, gravitationally bound system of point like masses collapses into black hole due to evaporation of energetic particles in binary gravitational collisions at the timescale $t_{evbin}=t_1 N/\ln N$ \cite{ZPod} or due collective effects at the timescale $t_{evcol}= t_1 N^{2/3}$ \cite{GurSav}, where $N \gg 1$ is the number of particles in the gravitationally bound system \cite{PBHrev}.

If gas of massive particles is dissipational, it evolution to black holes takes place at much smaller timescale, comparable with $t_1$. In particular, if magnetic monopole abundance is not suppressed by inflation and magnetic monopoles dominate in the Universe before their abundance is suppressed by monopole-antimonopole annihilation in gravitationally bound systems formed at the stage of their dominance, collapse into black holes turns out to be more rapid, than annihilation in these systems and magnetic monopole overproduction would convert into overproduction of PBHs \cite{PBHrev,kadnikov}. 

Inflation can end by sufficiently long MD stage of massive scalar field dominance, which can also result in PBH formation \cite{khlopov1}.

\subsubsection{PBH formation in first order phase transitions}
If inflation ends by first order phase transition or the symmetry breaking phase transition is a strong first order, the process of bubble nucleation can lead to black hole production in bubble wall collisions \cite{hawking3}. In the course of transition  bubbles of true vacuum, expanding in the false vacuum, collide and in the collision area the energy of bubble walls converts into a false vacuum bag, which separates from walls and pending on its mass either collapses in black hole \cite{kkrs1} or converts in oscillon \cite{oscilon}.

Bubble collisions become effective, when the bubble nucleation rate becomes equal to the rate of expansion, $H$, and the mass of forming black holes is determined by the energy of the false vacuum within a region with typical size of $1/H$.
\subsection{Primordial nonlinear structures}
Primordial objects created in the very early Universe seem to be constrained by the small size of cosmological horizon. However, inflation can provide large scale correlations in the space distribution of these objects, giving rise to the large scale primordial structures. 
\subsubsection{Archioles - large scale correlations of energy density of the axion-like fields}
In the axion-like models a complex scalar
field $\Psi = \psi \exp{(i \theta)}$ acquires after spontaneous symmetry breaking of global U(1) symmetry vacuum expectation value $\langle \psi \rangle = f$, leaving continuous degeneracy of vacua with arbitrary values of the phase $\theta$. This continuous degeneracy is broken by explicit
symmetry breaking term \begin{equation}
V_{eb} = \Lambda^{4} (1 -
\cos{\theta}).\label{ax} \end{equation} This term is negligible, if $f \gg \Lambda$.
In the axion models it doesn't exist at high temperature and appears due to instanton effects in the period of QCD phase transition. Then at $T \sim \Lambda$ takes place the second phase transition, in which continuous degeneracy of vacua is broken by the term Eq.(\ref{ax}) and the vacua have discrete degeneracy, corresponding to $\theta_{vac}= 0, 2 \pi, 4 \pi...$. The value of phase $\theta - \theta_{vac}$ acquires the meaning of the amplitude of axion field, which determines the energy density of the axion field oscillations. 

If first phase transition takes place after reheating, the continuous degeneracy of phase leads to singularities, having the geometric place of lines - axion strings.

After the second phase transition vacua with different values of $\theta_{vac}$ are separated by domain walls, surrounded by strings. This vacuum defect structure is unstable and rapidly decays, but the distribution of axion energy density follows the initial structure of walls-surrounded-by-strings. Since 80\% of axion string length corresponds to infinite strings, this structure provides large scale correlation in the distribution of axion energy density (see \cite{4} for review and references).
\subsubsection{Clusters of massive PBHs}
If the first phase transition takes place at the inflationary stage, the now observed part of the Universe acquires at the corresponding $e$-folding $N_i = 60$ unique value of phase $\theta_i$.

However, at successive steps of inflation with smaller $e$ foldings $N < N_i$ the value of phase experiences fluctuations $$\delta \theta \sim \frac{H_i}{2 \pi f},$$ where $H_i$ is the Hubble constant at the inflationary stage. Therefore, if $\theta_i< \pi$ at $N=N_i$ in some smaller regions fluctuations of $\theta$ can lead to values $\theta > \pi$. At successive stages of inflation with smaller $N$ fluctuations can lead in some smaller regions to the value of $\theta< \pi$. This process continues until the end of inflation.

In the result, at successive second phase transition, which takes place after reheating at $T\sim \Lambda \ll f$, the regions with $\theta < \pi$ and $\theta > \pi$ should be separated by closed domain walls. The process described above leads to a system of closed walls. Collapse of closed walls results in formation of black holes, which are not distributed stochastically but appear in clusters, in which black holes of smaller mass are created around the locally most massive black hole \cite{RubinCluster}.

This mechanism leads to formation of clusters of PBHs with masses, determined by the fundamental parameters of the model $f$ and $\Lambda$, which can have stellar and superstellar values. The minimal mass is determined by the condition that the width of domain wall ($\sim f/\lambda^2$) doesn't exceed the size of the gravitational radius of the wall. It gives \cite{AGN}
\begin{equation}
M_{min} = f(\frac{m_{Pl}}{\Lambda})^2.
\end{equation}

The principally maximal  mass of such PBHs is
determined by the condition that the wall does not dominate locally before it enters the cosmological horizon. Otherwise, local wall dominance leads to a superluminal $a \propto t^2$ expansion for the
corresponding region, separating it from the other part of the
universe. This~condition corresponds to the mass \cite{PBHrev}
\begin{equation}
M_{max} =
\frac{m_{Pl}}{f}m_{Pl}(\frac{m_{Pl}}{\Lambda})^2.\end{equation} 

Formation of PBHs in the collapse of closed walls is accompanied by the primordial gravitational wave (GW) background. Its spectrum is peaked at $$\nu_0=3 \times 10^{11}(\Lambda/f)\,{\rm Hz} $$ and the
energy density can be estimated as \cite{PBHrev} $\Omega_{GW} \approx 10^{-4}(f/m_{Pl})$. At $f \sim 10^{14}$ GeV this primordial gravitational wave background can reach $\Omega_{GW}\approx 10^{-9}.$ For the physically reasonable values of $1<\Lambda<10^8$~GeV the maximum of the spectrum corresponds to
\begin{equation}
3 \times 10^{-3}<\nu_0<3 \times  10^{5}\,{\rm Hz}.\end{equation}
This range may be within the reach of LIGO-VIRGO and future LISA detection of gravitational waves and this prediction may be of interest for interpretation of the recent results of the NANOGrav Collaboration \cite{nano}. 

The primordial origin of the observed massive and supermassive black holes \cite{dolgovPBH} may find additional support in the recent detection by LIGO and VIRGO collaborations of gravitational wave signal from a binary black hole merging with total mass $150 M_{odot}$ \cite{LV150}, corresponding to the gap in the predicted BH masses from first massive stars, can evidence for primordial origin of massive BHs \cite{LV150apl}.

\subsection{Antimatter stars as probes for nonhomogeneous baryosynthesis}
Any mechanism of baryosynthesis can under some conditions predict nonhomogeneous distribution of the baryon excess. In the extreme case, nonhomogeneity can lead not only to the spatial change of baryon asymmetry, but can also change its sign, so that antimatter excess can appear in some regions of baryon asymmetrical Universe \cite{CKSZ,DolgovAM,Dolgov2,Dolgov3,KRS2}.

Sufficiently large antimatter domains, corresponding to the mass, exceeding $10^3 M_{odot}$ can survive in the matter surrounding and form antimatter globular cluster in our Galaxy.
Owing to its situation in the galactic halo, where the gas density is low, and the absence of significant amount of matter gas within the cluster, $\gamma$ radiation from this cluster, can come dominantly from the surfaces of antimatter stars. It makes such object rather faint gamma source. Antimatter, lost by the cluster annihilates with the matter gas and is the source of gamma background. It puts upper limit on the mass of cluster around $10^5 M_{odot}$.

Antimatter supernova explosions can accelerate antinuclei and generate heavy antinuclear component of cosmic rays. Since the estimated flux of secondary cosmic antihelium, originated from cosmic ray interaction with matter, is far beyond the sensitivity of AMS02 experiment, detection of antihelium in this experiment would be a very strong evidence for its primordial nature and for existence of antimatter stars in our Galaxy \cite{AMS}. It may provide distinction of this mechanisms from other predictions of possible forms of antimatter in our Galaxy \cite{Blinnikov}.

The first claims on the detection of antihelium events in the AMS02 experiment can hardly find explanation by natural astophysical sources \cite{poulin} and, if confirmed in the future data analysis, may strongly evidence for the existence of antimatter stars in our Galaxy.

\section{Conclusion}
Multimessenger cosmology of new physics deals with hypothetical phenomena, which are not predicted with necessity in the framework of the now standard cosmological paradigm. Being model dependent cosmological consequences of BSM physics, their signatures can specify the underlying particle models and provide their effective selection. Positive results of the searches for such signatures would lead to nonstandard deviations from the modern cosmological standards, specifying true cosmological scenario and fundamental structure of the microworld, on which it is based.

\section*{Acknowledgement} I express my gratitude to MDPI Journal "Physics" for information support. The work was supported by grant of the Russian Science Foundation (Project No-18-12-00213).


\section*{References}
\begin{thebibliography}{99}
\bibitem{Lindebook}
  Linde A D  1990 {\it Particle Physics and Inflationary Cosmology} (Chur:Harwood).
\bibitem{Kolbbook}
 Kolb E W and Turner M S  1990 {\it The Early Universe} 
(Boston, MA,:Addison-Wesley).
\bibitem{Rubakovbook1}
Gorbunov  D S and Rubakov  V A 2011 {\it Introduction to the Theory of the Early Universe Hot Big Bang Theory. Cosmological Perturbations and Inflationary Theory}(Singapore: World~Scientific).
\bibitem{Rubakovbook2}
Gorbunov  D S and Rubakov  V A 2011 {\it Introduction to the Theory of the Early Universe Hot Big Bang Theory} (Singapore: World~Scientific).
\bibitem{book}   Khlopov M Y 1999 {\it Cosmoparticle Physics} (Singapore: World~Scientific).
\bibitem{newBook}
 Khlopov M Y  2012
{\em Fundamentals of Cosmoparticle Physics} (Cambridge, UK: CISP-Springer).
\bibitem{4} Khlopov M 2016 {\it Symmetry} {\bf 8} 81. 

\bibitem{DMRev}   Khlopov M 2013 {\it Int. J.  Mod. Phys.}  A {\bf 28} 1330042.  

\bibitem{ijmpd19} Khlopov  M Yu 2019 {\it Int. J.  Mod. Phys.} D {\bf 28} 1941012.
\bibitem{ketovSym} Ketov S V, Khlopov M Yu 2019 {\it Symmetry} {\bf 11} 511. 
\bibitem{PBHrev}
 Khlopov M Y 2010  {\it Res. Astron. Astrophys.}{\bf 10} 495.
\bibitem{ZN}  Zeldovich Y B,  Novikov I D 1967 {\it Sov. Astron.} {\bf 10} 602. 
\bibitem{carr75}  Carr B J 1975 {\it Astroph. J.} {\bf 201} 1.
\bibitem{polnarev}  Polnarev A G,  Khlopov M Y 1985 {\it Sov. Phys. Uspekhi} {\bf 28} 213.  
\bibitem{khlopov0}
 Khlopov M Y,  Polnarev A G 1980 {\it Phys. Lett.} B {\bf 97} 383.
\bibitem{ZPod}  Zeldovich Y B ,  Podurets M A 1965 {\it Sov. Astron.} {\bf 9} 742. 
\bibitem{GurSav}  Gurzadian V G,  Savvidi G K 1986 {\it Astrophys. J.} {\bf 160}  203. 
\bibitem{kadnikov} Kadnikov A F, Khlopov M Y  and Maslyankin V I 1990 {\it Astrophysics} {\bf 31} 523.
 \bibitem{khlopov1}
 Khlopov M Y,  Malomed B A ,  Zel'dovich Y B 1985 {\it Mon. Not. R. Astron. Soc.} {\bf 215} 575.
 \bibitem{hawking3}
 Hawking S W ,  Moss I G,  Stewart J M 1982 {\bf Phys. Rev.} D {\bf 26} 2681.
 \bibitem{kkrs1} Konoplich  R V,  Rubin S G , Sakharov A S, Khlopov M Y 1998 {\it Astron. Lett.} {\bf 24} 413. 
 \bibitem{oscilon}  Dymnikova I, Khlopov M Y, Koziel  L and  Rubin S  G 2000 {\it Gravitation and Cosmology} {\bf 6}, 311.
\bibitem{AGN}    Rubin S G ,  Sakharov A S, Khlopov M Y 2001 {\it JETP} {\bf 92} 921.
\bibitem{RubinCluster}  Belotsky K M, Dokuchaev V I, Eroshenko Y N,  Esipova E A,  Khlopov M Y, Khromykh L A, Kirillov A A, Nikulin V V, Rubin S G, Svadkovsky I V 2019 {\it Eur. Phys. J.} C {\bf 79} 246.
  M.Y.; Khromykh, L.A.; Kirillov,~A.A.; Nikulin, V.V.; Rubin, S.G.; 
\bibitem{nano}  Arzoumanian Z {\it et al} [NANOGrav Collaboration] 2020  The NANOGrav 12.5-year Data Set: Search For An Isotropic Stochastic Gravitational-Wave Background,{\it Preprint}  arXiv:2009.04496.
\bibitem{dolgovPBH} Dolgov A D 2018 {\it Int.J.Mod.Phys.} A {\bf 31} 1844029.
\bibitem{LV150} The LIGO Scientific Collaboration; the Virgo Collaboration; Abbott R {\it et al} 2020 {\it Phys. Rev. Lett.} {\bf 125} 101102.
\bibitem{LV150apl} The LIGO Scientific Collaboration; the Virgo Collaboration; Abbott R {\it et al} 2020 {it Astrophys J. Lett.} {\bf 900} L13.
\bibitem{CKSZ} Chechetkin V M, Khlopov M Yu,  Sapozhnikov M G,  Zeldovich Y B 1982 {\it Phys. Lett.} B {\bf 118} 329.  
 \bibitem{DolgovAM}    Dolgov A D 2002 {\it Nucl. Phys. Proc. Suppl.} {\bf 113} 40.
\bibitem{Dolgov2}  Dolgov A, Silk J 1993 {\it Phys. Rev.} D {\bf 47} 4244. 
\bibitem{Dolgov3} Dolgov A D, Kawasaki  M,  Kevlishvili N 2009 {\it Nucl. Phys.}  B {\bf 807} 229.
\bibitem{KRS2}   Khlopov  M Y, Rubin S G,  Sakharov A S 2000 {\it Phys. Rev.} D {\bf 62} 083505.  
\bibitem{AMS}
Belotsky K M, Golubkov  Y A, Khlopov M Y, Konoplich R V, Sakharov A S 2000 {\it Phys. Atom. Nucl.} {\bf 63} 233. 

\bibitem{Blinnikov}
  Blinnikov S I,  Dolgov A D,  Postnov K A 2015 {\it Phys. Rev.} D {\bf 92} 023516.   
\bibitem{poulin}  Poulin V, Salati P,  Cholis I, Kamionkowski M, Silk J 2019 {\it Phys. Rev.} D {\bf 99} 023016. 
\end{thebibliography}
\end{document}